\documentclass{article}
\usepackage{spconf}
\usepackage{color}
\usepackage{bm}
\usepackage{bbm}
\usepackage{amsmath}
\usepackage{amssymb}
\usepackage{amsthm}
\usepackage{graphicx}
\usepackage{mathrsfs}
\usepackage{empheq}	
\usepackage[mathcal]{euscript}
\usepackage[margin = 2cm]{geometry}
\graphicspath{{figs/}}
\usepackage{lipsum}
\usepackage{cite}
\usepackage{url}
\usepackage{balance}
\newtheorem{theorem}{Theorem}

\newtheorem{lemma}{Lemma}

\renewcommand{\P}{\mathbb{P}}
\newcommand{\E}{\mathbb{E}}
\newcommand{\beq}{\begin{equation}}
\newcommand{\eeq}{\end{equation}}
\newcommand{\beqa}{\begin{eqnarray}}
\newcommand{\eeqa}{\end{eqnarray}}

\newcommand{\T}{^\top}

\DeclareMathOperator*{\argmin}{arg\,min}
\newcommand*{\QED}{\hfill\ensuremath{\blacksquare}}

\title{Network Classifiers Based on Social Learning}
\name{Virginia Bordignon$^{\star}$\qquad Stefan Vlaski$^{\star}$\qquad Vincenzo  Matta$^{\dagger}$ \qquad Ali H. Sayed$^\star$\thanks{This work was supported in part by the Swiss National Science Founda- tion grant 205121-184999.
		E-mails: virginia.bordignon@epfl.ch, stefan.vlaski@epfl.ch, vmatta@unisa.it, ali.sayed@epfl.ch.}}

\address{$^{\star}$ School of Engineering, Ecole Polytechnique F\'ed\'erale de Lausanne (EPFL)\\$^{\dagger}$DIEM, University of Salerno}
\begin{document}
\ninept
\maketitle
\begin{abstract}
This work proposes a new way of combining independently trained classifiers over space and time. Combination over space means that the outputs of {\em spatially distributed} classifiers are aggregated. {Combination over time means that the classifiers respond to streaming data during testing and continue to improve their performance even during this phase.} {By doing so, the proposed architecture is able to improve prediction performance over time with unlabeled data.} Inspired by social learning algorithms, which require prior knowledge of the observations distribution, we propose a Social Machine Learning (SML) paradigm that is able to exploit the imperfect models generated during the learning phase. We show that this strategy results in consistent learning with high probability, and it yields a robust structure against poorly trained classifiers. Simulations with an ensemble of feedforward neural networks are provided to illustrate the theoretical results.
\end{abstract}

\vspace{-5pt}
\section{Introduction and Related Work}
\vspace{-5pt}
\label{sec:intro}
{Social learning strategies allow the classification of unlabeled features by a {\em heterogeneous} network of agents~\cite{jadbabaie2012non,krishnamurthy2013social,nedic2017fast,salami2017social,lalitha2018social,matta2020interplay,bordignon2020adaptive,zhao2012learning}. The heterogeneity of the network is twofold: first, agents may be observing different (possibly non-overlapping) sets of attributes of the same underlying phenomenon; second, their statistical models need not be the same, e.g., two agents may be observing the same attribute from different perspectives. Neighboring agents share statistics about the observed features and diffuse this information across the network to arrive at a conclusion on the nature of the observed phenomenon.}

Many social learning approaches exist in the literature that have been shown to yield correct asymptotic learning of the true state of nature under mild identifiability assumptions~\cite{jadbabaie2012non,nedic2017fast,salami2017social,lalitha2018social,bordignon2020adaptive,zhao2012learning}. These results, however, come at a cost: the strategies require prior knowledge of the true underlying distributions for the features. { In practice we often have access to feature data only, or to some approximate models for the distributions.} For example, uncertain likelihoods in social learning have been considered in~\cite{hare2020non}, albeit only for multinomial distributions. In this work, we will allow for a fairly broad class of distributions.

Another distinguishing aspect is that we will consider cooperation among {\em spatially distributed} classifiers, and aggregation over {\em time} of the inference produced from streaming data. An ensemble of classifiers is known to be a more robust structure than an isolated, perhaps poorly trained, classifier \cite{kittler1998combining}. Examples of ensemble approaches are Bagging \cite{breiman1996bagging} and Boosting \cite{freund1997decision}, in which classifiers combine weighted decisions across \emph{space}. Boosting requires labeled samples to tune the combinations weights. Both bagging and boosting methods do not address the streaming data case. {Other examples include localized Gaussian Process Regression (GPR) methods~\cite{tresp2000bayesian, nguyen2008local,lederer2020real}, which require labeled samples for online training and focus on kernel-based classifiers. In our work, we are interested in more general classifier structures.}

{We therefore propose the Social Machine Learning (SML) approach: a decentralized algorithm for combining the outputs of a heterogeneous network of classifiers over space and time, based on the adaptive diffusion algorithm proposed in \cite{matta2018estimation,bordignon2020adaptation,bordignon2020adaptive}. The SML structure inherits the following qualities from social learning: the ability to combine classifiers with different dimensions and statistical models, while providing asymptotic performance guarantees in addition to continuous performance improvement even during the prediction phase.}
We show that $i)$ with high probability, consistent learning occurs despite the imperfectly trained models; and $ii)$ poorly trained classifiers can leverage the networked setup to improve their performance. We exploit these results particularly for the setup of a network of feedforward neural networks (FNN).

\emph{Notation: }Random variables are written in bold font and deterministic variables in normal font.  $\E_x(\cdot)$ and $\P_x(\cdot)$ respectively denote the expectation and probability measure computed with respect to the single random variable $\bm{x}$. 

\vspace{-6pt}
\section{The Decision-Making Problem}
\vspace{-5pt}
A network of $K$ agents is engaged to accomplish the following decision-making task. There is a {\em true} underlying binary state of nature represented by an {equiprobable binary random variable $\bm{\gamma}\in\{-1,+1\}\triangleq \Gamma$}.
As time progresses, each agent collects streaming data arising from the true state of nature. More specifically, agent $k=1,2,\ldots,K$ observes at times $i=1,2,\ldots,$ the random \emph{feature vectors} $\bm{h}_{k,i}\in\mathcal{H}_k$, which are independent and identically distributed (i.i.d.) over time (but {\em not} necessarily across the agents). 
The features $\bm{h}_{k,i}$ at agent $k$, given the underlying true hypothesis $\gamma$, form a sequence of i.i.d. random vectors distributed according to some conditional distribution (or likelihood):
\begin{equation}
\bm{h}_{k,i}\sim L_k(h|\gamma),~~h\in\mathcal{H}_k,\gamma\in\Gamma.\label{eq:like}
\end{equation}
{We allow the feature vectors to have different dimensions and attributes across the agents.} The goal of the decision learning task is to let each agent learn, as $i\rightarrow\infty$, the right hypothesis $\gamma$. 

If agent $k$ knows the true joint distribution of features and label, it can then apply the paradigm of \emph{Bayes classifiers}~\cite{mohri2018foundations}. {The Bayes classifier is the solution to a maximum-a-posteriori (MAP) problem, where the label estimated by classifier $k$ is the label $\gamma$ that maximizes $p_k(\gamma|\bm{h}_{k,1},\bm{h}_{k,2},\dots, \bm{h}_{k,i})$, i.e., the posterior probability of $\gamma$ given the sequence of features $\{\bm{h}_{k,j}\}_{j=1}^i$.} However, agents might not have enough information to solve this classification problem alone, e.g., if the signals at agent $k$ are not informative enough (for example, it may be the case that $L_k(h|-1)=L_k(h|+1)$ for all $h\in\mathcal{H}_k$). If however the network as a whole possesses enough information, under the weaker assumption of global identifiability, then a \emph{social learning} scheme can be used and allows agents to learn the truth \cite{jadbabaie2012non,zhao2012learning,salami2017social,lalitha2018social, bordignon2020adaptive}. 

In practice, the statistical characterization \eqref{eq:like} of the features and/or labels is often unknown. We will see next how the individual classifiers can be trained to approximate the unknown distributions.

\vspace{-5pt}
\section{Local Instantaneous Classifiers}
\vspace{-5pt}
The fundamental assumption of this work is that the likelihoods $L_k(h|\gamma)$ are {\em unknown}. {To circumvent this lack of knowledge, we assume that each agent is able to train locally some standard binary classifier during a \emph{training} phase. In order to avoid confusion, the random variables pertaining to the training set are topped with a sign $\sim$. Whenever we are dealing with the training phase of the classifiers, feature vectors and labels are indexed with the time subscript $n$. For the \emph{prediction} (i.e., testing)  phase, we use the time subscript $i$. }

Agent $k$ is trained by collecting $N_k$ examples constituted by pairs $\{\widetilde{\bm{h}}_{k,n},\widetilde{\bm{\gamma}}_n\}_{n=1}^{N_k}$. Labels $\widetilde{\bm{\gamma}}_n$ are uniformly distributed over $\Gamma=\{+1,-1\}$ so the pair $(\widetilde{\bm{h}}_{k,n},\widetilde{\bm{\gamma}}_n)$ is distributed according to the joint distribution:
{\begin{equation}
p_k(h,\gamma)=p_k(\gamma)L_k(h|\gamma),~~h\in\mathcal{H}_k,~\gamma\in\Gamma,
\end{equation}
with the uniform prior $p_k(\gamma)=1/2$ for $\gamma\in\Gamma$. We are interested in the following statistic:
\begin{equation}
\log\frac{p_k(+1|\bm{h}_{k,i})}{p_k(-1|\bm{h}_{k,i})}\stackrel{\text{(a)}}{=}\log\frac{L_k(\bm{h}_{k,i}|+1)}{L_k(\bm{h}_{k,i}|-1)}\label{eq:postratio}
\end{equation}
where the equality in (a) follows from the Bayes rule and the uniform priors assumption. The log-likelihood ratio on the RHS of \eqref{eq:postratio} is positive whenever the observation $\bm{h}_{k,i}$ is more likely to have come from class $+1$ and negative when it is more likely to have originated from class $-1$. This is the same sufficient statistic aggregated over space and time in social learning~\cite{bordignon2020adaptation,bordignon2020adaptive} and in signal detection schemes~\cite{matta2018estimation,kay1993fundamentals,poor2013introduction}. 

After training, the classifier will generate approximate posterior models $\widehat{p}_k(\gamma|h)$. Thus, instead of \eqref{eq:postratio}, we will rely on the following \emph{logit} statistic:
\begin{align}
\log\frac{\widehat{p}_k(+1|\bm{h}_{k,i})}{\widehat{p}_k(-1|\bm{h}_{k,i})}=\log\frac{\widehat{p}_k(+1|\bm{h}_{k,i})}{1-\widehat{p}_k(+1|\bm{h}_{k,i})}\triangleq f_k(\bm{h}_{k,i}).\label{eq:logit}
\end{align}
The function $f_k$ belongs to a specific class of functions $\mathcal{F}_{k}:\mathcal{H}_k\mapsto \mathbb{R}$ that depends on the choice of classifier. For example, in logistic regression with $h\in\mathbb{R}^M$, $\mathcal{F}_{k}$ is parameterized by a vector $w\in\mathbb{R}^M$, and we have the linear logit function
$f_k(h;w)=w\T h$~\cite{mohri2018foundations}. Since the logit in \eqref{eq:logit} operates on the feature vector collected by an {\em individual} agent in a {\em single time instant}, we will refer to \eqref{eq:logit} as a {\em local instantaneous} logit. }

The local instantaneous classifiers are trained by choosing the function $f$ within $\mathcal{F}_k$ that minimizes a suitable risk function $R_k(f)$. We define this optimal function as the \emph{target model}:
\begin{equation}
f^o_k\triangleq \argmin_{f\in\mathcal{F}_k} R_k(f).\label{eq:w0}
\end{equation}
In this work we focus on the logistic risk:
\begin{equation}
R_k(f)=\E_{h_{k},\gamma}
\log
\left(1+e^{-\widetilde{\bm{\gamma}}_nf(\widetilde{\bm{h}}_{k,n})}\right),
\label{eq:exprisk}
\end{equation}
{which is commonly used for binary classification tasks for traditional classifiers such as logistic regression or more complex structures such as neural networks with softmax output layers.  }
Note that the expectation is computed under the (unknown) joint distribution of the pair $(\widetilde{\bm{h}}_{k,n},\widetilde{\bm{\gamma}})$. 
Since all agents rely on a finite set of training samples, they will solve instead an empirical optimization problem:
\begin{equation}
\widetilde{\bm{f}}^N_{k}\triangleq \argmin_{f\in\mathcal{F}_k}\widetilde{\bm{R}}^N_{k}(f),\label{eq:emprisk}
\end{equation}
where the empirical risk is in the form of an empirical logistic risk:
\begin{equation}
\widetilde{\bm{R}}^N_{k}(f)=\frac{1}{N_{k}}\sum_{n=1}^{N_{k}}\log
\left(1+e^{-\widetilde{\bm{\gamma}}_nf(\widetilde{\bm{h}}_{k,n})}\right),
\end{equation}
which is computed over the training set. Since agents solve \eqref{eq:emprisk} until convergence, we assume they reach an empirical minimizer $\widetilde{\bm{f}}^N_{k}$ that is close enough to the target $f_k^o$ for sufficiently large $N_k$ under ergodicity assumptions. Next, we introduce the algorithm that allows these classifiers to be combined within a network.

\vspace{-5pt}
\section{Social Machine Learning}
\begin{figure}[tb]
	\centering
	\includegraphics[width=.9\linewidth]{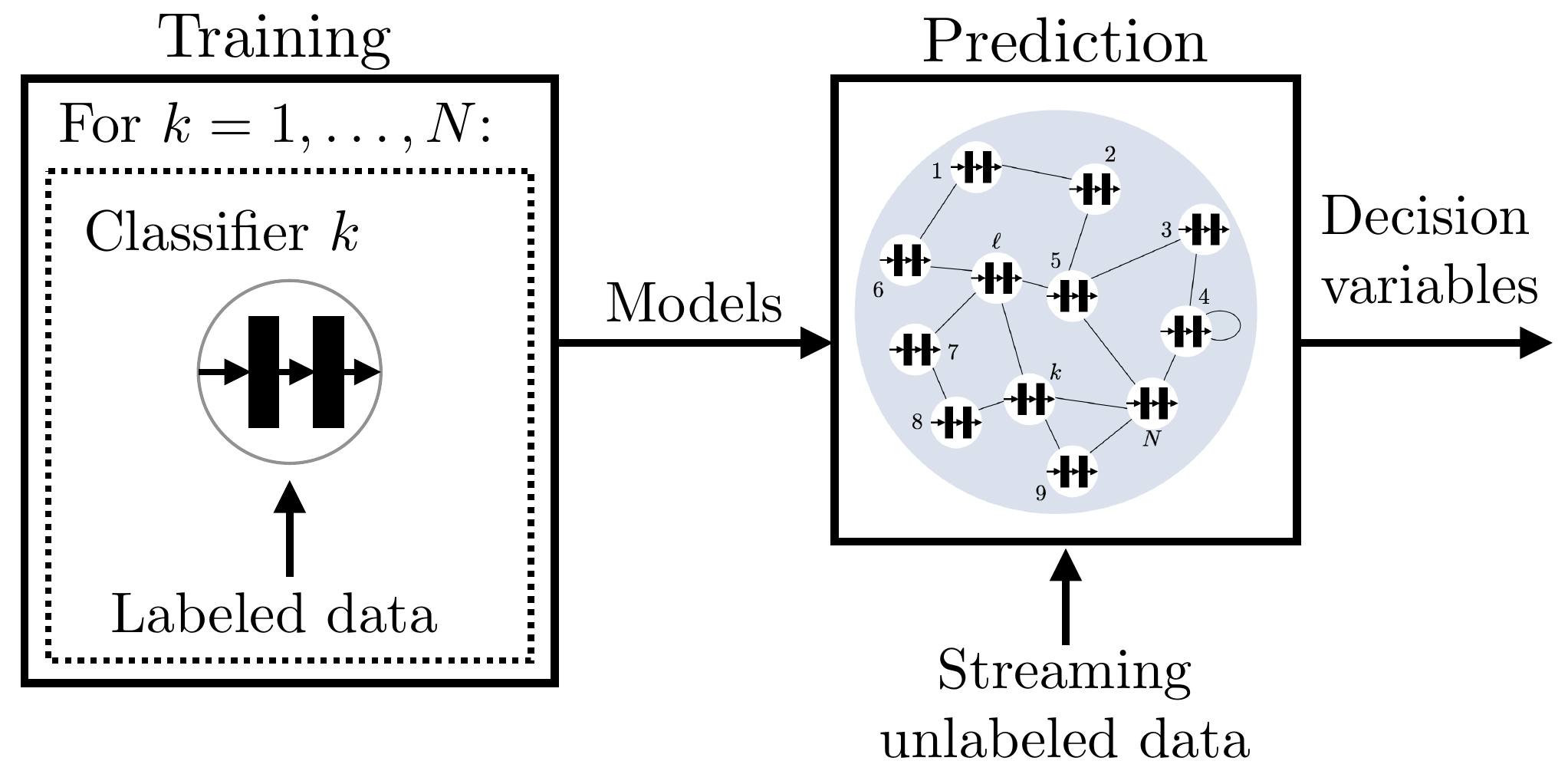}\vspace{-10pt}
	\caption{{Social Machine Learning (SML) diagram.}}\label{fig:diag}\vspace{-15pt}
\end{figure}

\vspace{-5pt}
The network is modeled as a strongly connected graph (i.e., where there is always a path in both directions between any two agents and at least one self-loop) with a left-stochastic combination matrix $A$, whose elements $a_{\ell k}$ are nonnegative, and $a_{\ell k}=0$ if agent $\ell\notin \mathcal{N}_k$, where $\mathcal{N}_k$ denotes the neighborhood of agent $k$. Under this condition, we define the Perron eigenvector $\pi$ as \cite{sayed2014adaptation}:
\begin{equation}
A\pi=\pi,~~\textstyle\sum_{k=1}^K\pi_k=1,~~\pi_k>0, \text{ for all }k=1,2,\dots,K.
\end{equation}
During the prediction phase, agents are observing {\em unlabeled} streaming private features $\bm{h}_{k,i}$. {In Fig.~\ref{fig:diag}, we show a diagram depicting the SML approach. After the training phase, the posterior models are used in the prediction phase to form the individual agent's decisions in a social learning setup.}

In~\cite{bordignon2020adaptive,bordignon2020adaptation}, an adaptive version of social learning was introduced, where agents update their beliefs (or opinions) $\bm{\varphi}_{k,i}(\gamma)$ as\footnote{The belief $\bm{\varphi}_{k,i}(\gamma)$ quantifies the confidence of agent $k$ at instant $i$ that $\gamma$ is the true state of nature.}:
\begin{align}
\bm{\psi}_{k,i}(\gamma)&=\frac{\bm{\varphi}_k^{1-\delta}(\gamma)L_k^{\delta}(\bm{h}_{k,i}|\gamma)}{\sum_{\gamma'\in\Gamma}\bm{\varphi}_k^{1-\delta}(\gamma')L_k^{\delta}(\bm{h}_{k,i}|\gamma')}\label{eq:bayes}\\
\bm{\varphi}_{k,i}(\gamma)&=\frac{\exp\left\{\sum_{\ell=1}^{K}a_{\ell k}\log \bm{\psi}_{\ell,i}(\gamma)\right\}}{\sum_{\gamma'\in\Gamma}\exp\left\{\sum_{\ell=1}^{K}a_{\ell k}\log \bm{\psi}_{\ell,i}(\gamma')\right\}}\label{eq:comb}
\end{align}
where $0<\delta\ll 1$ is a small step-size parameter. In \eqref{eq:bayes}, the agent uses its private observation $\bm{h}_{k,i}$ to update its belief into an intermediate belief $\bm{\psi}_{k,i}(\gamma)$. In \eqref{eq:comb}, the agent combines the intermediate beliefs coming from neighbors into its updated belief $\bm{\varphi}_{k,i}(\gamma)$.  {Note that these relations rely on knowledge of the exact likelihood functions $L_{k}(h|\gamma)$, whereas in this work these likelihoods will be estimated during the training phase. The objective is to show that with minimal pre-training, the estimated likelihoods will enable the social learning algorithm to classify unlabeled data correctly with high probability and, moreover, the confidence of the classifier in its decisions will continually grow over time in response to streaming data. This property is fundamentally different from existing static testing phases for traditional classifiers, where classification decisions are instantaneous and are not exploited to improve performance. 

An equivalent way of representing \eqref{eq:bayes} and \eqref{eq:comb} is in the form of an \emph{adaptive diffusion strategy}:
\beq
\bm{\lambda}_{k,i}=(1-\delta)\sum_{\ell=1}^K a_{\ell k} \bm{\lambda}_{\ell,i-1} + 
\delta \sum_{\ell=1}^K a_{\ell k} \bm{c}_{\ell,i},
\label{eq:diffstat}
\eeq 
where we defined $\bm{\lambda}_{k,i}\triangleq\log [\bm{\varphi}_{k,i}(+1)/\bm{\varphi}_{k,i}(-1)]$ and $\bm{c}_{k,i}$ is taken as the log-likelihood ratio seen in the RHS of~\eqref{eq:postratio}. {Eq.~\eqref{eq:diffstat} has moreover the form of a distributed stochastic
gradient algorithm with step-size $\delta$ and with a quadratic cost function -- see \cite{matta2018estimation,sayed2014adaptation}.} The algorithm in~\eqref{eq:diffstat} constructs the aggregate classification variable $\bm{\lambda}_{k,i}$ from the past information, in the shape of $\bm{\lambda}_{k,i-1}$, and the present information $\bm{c}_{k,i}$ received from neighboring classifiers. The structure in \eqref{eq:diffstat} will enable cooperation over space and time. }

In our approach, in the place of $\bm{c}_{k,i}$, we will consider the local instantaneous logit statistic $f_k(\bm{h}_{k,i})$.
We also assume that each agent performs a \emph{debiasing} operation before sharing the statistic $f_k(\bm{h}_{k,i})$, by discounting its empirical mean over the training dataset. We define this \emph{empirical training mean} as:
\begin{equation}
\widetilde{\bm{\mu}}^N_{k}(f_k)=\frac{1}{N_k}\sum_{n=1}^{N_k}f_k(\widetilde{\bm{h}}_{k,n}).
\end{equation}
The diffusion strategy in \eqref{eq:diffstat} is then run with the choice:
\begin{equation}
	\bm{c}_{k,i}=f_k(\bm{h}_{k,i})-\widetilde{\bm{\mu}}^N_{k}(f_k)
\label{eq:choiceofc}
\end{equation}
Note that $\bm{c}_{k,i}$ contains two independent sources of  randomness. The first, introduced by $f_k(\bm{h}_{k,i})$, contains the randomness from the prediction sample $\bm{h}_{k,i}$. The second source of randomness comes from the training samples $\widetilde{\bm{h}}_{k,n}$, which are introduced in the term $\widetilde{\bm{\mu}}^N_{k}(f_k)$. The prediction and training feature vector samples are independent of each other. 

The instantaneous decision of agent $k$, namely $\widehat{\bm{\gamma}}_{k,i}$, is taken according to the rule:
\begin{equation}
\widehat{\bm{\gamma}}_{k,i}=\text{sign}\left(\bm{\lambda}_{k,i}\right)\label{eq:decirule},
\end{equation}
where $\text{sign}(x)=+1$, if $x\geq 0$ and $\text{sign}(x)=-1$ otherwise. This choice is motivated by the fact that the logarithmic ratios in \eqref{eq:logit} are positive whenever the fiducial posterior probability $\widehat{p}_k(+1|h)$ exceeds $1/2$, and are negative otherwise.

From previous work \cite{bordignon2020adaptive} we know that, for sufficiently small values of the step-size $\delta$, the adaptive diffusion strategy in \eqref{eq:diffstat} with decision rule in \eqref{eq:decirule} is able to learn consistently\footnote{In our setting, consistent learning means that the classification error probability can be made arbitrarily small by suitably reducing the value of the step-size $\delta$.} the true hypothesis under the following condition. Let
\begin{align}
\mu^+_k(f_k)\triangleq \E_{L_k(h|+1)}f_{k}(\bm{h}_{k,i}),\quad
\mu^-_k(f_k)\triangleq \E_{L_k(h|-1)}f_{k}(\bm{h}_{k,i}),
\end{align}
where the notation $\E_{L_k(h|\gamma)}$ indicates that the expectation is computed under the distribution $L_k(h|\gamma)$. Let also
\begin{equation}
\mu^{+}(f)\triangleq
\sum_{k=1}^K \pi_{k}\mu^+_k(f_k),\quad
\mu^{-}(f)\triangleq
\sum_{k=1}^K \pi_{k}\mu^-_k(f_k),
\end{equation}
where we use the compact notation $f$ to indicate the dependency of the above averages on the group of functions $f_1,f_2,\dots,f_K$. Then, consistent learning is achieved if:
\begin{equation}
\mu^+(f)>\widetilde{\bm{\mu}}^{N}(f)~~\text{ and }~~\mu^-(f)<\widetilde{\bm{\mu}}^N(f).
\label{eq:log2exprisk}
\end{equation}
For each agent, the result of the training phase is the optimal empirical classifier function $\widetilde{\bm{f}}_k^N$. Therefore, we are interested in determining if both events described in \eqref{eq:log2exprisk} are likely to simultaneously occur when the classifier functions are given by $\widetilde{\bm{f}}^N$, i.e., by the group of functions $\widetilde{\bm{f}}_1^N, \widetilde{\bm{f}}_2^N,\ldots,\widetilde{\bm{f}}_K^N$.
\begin{figure*}[ht]
	\centering
	\includegraphics[width=.23\linewidth]{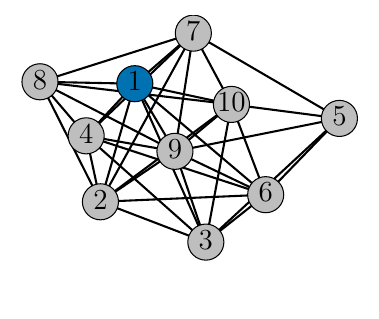} \hspace{10pt}
	\includegraphics[width=.31\linewidth]{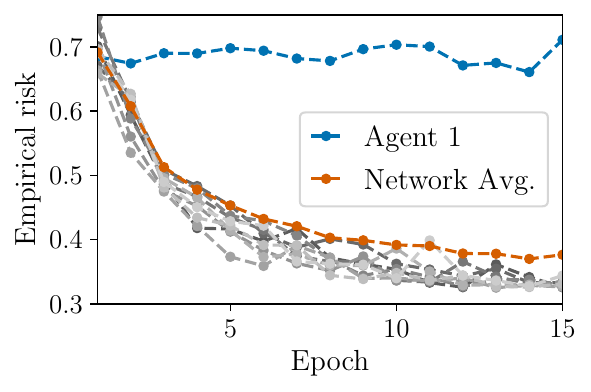}\hspace{10pt}
	\includegraphics[width=.3101\linewidth]{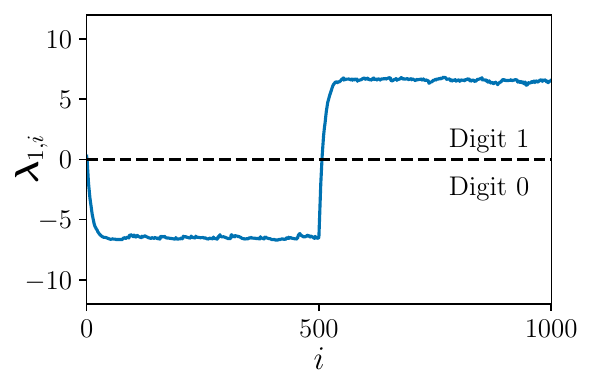}\vspace*{-10pt}
	\caption{{Network classifiers for handwritten digits classification. \emph{Leftmost panel}: Network topology. \emph{Middle panel}: Empirical risk evolution for agent 1 (in blue), the rest of the agents (different shades of gray) and for the network average empirical risk (in red). \emph{Rightmost panel}: Decision variable of agent $1$ over the prediction phase, where the dashed line indicates the decision boundary between digits 1 and 0.}}
	\label{fig:net} \vspace{-15pt}
\end{figure*}

\subsection{SML Consistency}
In Theorem~\ref{the:consist}, we will show that the SML strategy consistently learns the truth with high probability, as the number of training samples grows and for a moderately complex classifier structure. 
The complexity of the classifier structure is related to the complexity of the class of functions $\mathcal{F}_k$. The latter is quantified by using the concept of \emph{Rademacher average} (initially introduced as Rademacher penalty in \cite{koltchinskii2001rademacher}). We follow the definition in \cite{boucheron2005theory} and introduce, for a class of functions $\mathcal{F}$ and $N$ samples $x_1,x_2\dots,x_N\in\mathcal{X}$, the set of vectors $\mathcal{F}\left(x_1^N\right)$ as $(f(x_1),f(x_2),\dots,f(x_N))$ with $f\in\mathcal{F}$. Then, the (empirical) Rademacher average associated with $\mathcal{F}\left(x_1^N\right)$ is:
\begin{equation}
\mathcal{R}\left(\mathcal{F}\left(x_1^N\right)\right)\triangleq\E_{r}\left|\sup_{f\in \mathcal{F}}\frac{1}{N}\sum_{n=1}^N\bm{r}_nf(x_n)\right|,\label{eq:defradem}
\end{equation}
where $\bm{r}_n$ are independent and identically distributed Rademacher random variables, i.e., with $\P(\bm{r}_n=1)=\P(\bm{r}_n=-1)=1/2$. 

\begin{theorem}[\textbf{SML Consistency}]\label{the:consist} For the logistic loss, assume that $R(f^o)<\log 2$ and that $f_k(h_k)<B$ for every $h_k\in\mathcal{H}_k$ and $k=1,2,\dots,K$, with $B>0$. For any $d\in (0, -\log(e^{R(f^o)}-1))$, we have the following bound for the probability of consistent learning:
	\begin{align}
	&\P\left(\mu^+(\widetilde{\bm{f}}^N)>\widetilde{\bm{\mu}}^{N}(\widetilde{\bm{f}}^N)\,,\,\mu^-(\widetilde{\bm{f}}^N)<\widetilde{\bm{\mu}}^N(\widetilde{\bm{f}}^N)\right)\nonumber\\
	&\geq 1-2 \sum_{k=1}^K\exp\left\{\frac{-\left(d-\rho^{(k)}_N\right)^2N_k}{2B^2} \right\}\nonumber\\
	&-\sum_{k=1}^K \exp\left\{\frac{-\left(\frac{\Delta -R(f^o)}{2}-\rho^{(k)}_N\right)^2N_k}{2B^2} \right\},\label{eq:theocons}
	\end{align}
	with $\Delta\triangleq\log (1+e^{-d})$, $R(f^o)=\sum_{k=1}^K\pi_kR_k(f_k^o)$ and
	\begin{equation}
	\rho^{(k)}_N\triangleq  2 \E_{h_k}\mathcal{R}(\mathcal{F}_k({\bm{h}}_{1}^{N_k})).\label{eq:rho}
	\end{equation}
	\QED
\end{theorem}\vspace{-5pt}
\emph{Sketch of proof:} The proof cannot be included for space limitations, but we present some insights for it. First, define the average network risk as $R(f)\triangleq \sum_{k=1}^K\pi_kR_k(f_k)$. 
We have that:
\begin{align}
&R(\widetilde{\bm{f}}^N)\stackrel{\text{(a)}}{\geq} \sum_{k=1}^K\pi_k\log \Big( 1+\exp\Big(-\E_{h_k,\gamma}\bm{\gamma}\widetilde{\bm{f}}^N_k(\bm{h}_{k,i}) \Big)\Big)\nonumber\\
&\stackrel{\text{(b)}}{\geq} \log \Big( 1+\exp\Big(-\sum_{k=1}^K\pi_k\E_{h_k,\gamma}\bm{\gamma}\widetilde{\bm{f}}^N_k(\bm{h}_{k,i})\Big)\Big)\nonumber\\
&=\log \Bigg( 1+\exp\Bigg(-\frac{(\mu^+(\widetilde{\bm{f}}^N)-\mu^-(\widetilde{\bm{f}}^N)) }{2}\Bigg)\Bigg),\label{eq:ineq}
\end{align}
where in (a) and (b) we used Jensen's inequality with the convexity of $\log(1+e^x)$. The inequality in~\eqref{eq:ineq} translates into:
\begin{align}
R(\widetilde{\bm{f}}_k^N)\leq  \log\hspace{-1pt}\left(\hspace{-1pt}1+e^{-d} \hspace{-1pt}\right)\hspace{-2pt}\implies\hspace{-2pt} \frac{\mu^+(\widetilde{\bm{f}}^N)-\mu^-(\widetilde{\bm{f}}^N)}{2}\hspace{-2pt}\geq \hspace{-2pt}d.\label{eq:ineq2}
\end{align}
If now the risk in~\eqref{eq:ineq2} is sufficiently close to the risk $R(f^o)<\log 2$, we see that $d>0$. In other words, a good generalization capability of the trained classifiers $\widetilde{\bm{f}}_k^N$, i.e., a lower risk, implies a larger gap between means $\mu^+$ and $\mu^-$. In view of~\eqref{eq:log2exprisk}, this gap can be sufficient to achieve consistent learning, provided that the concentration error of the empirical training mean around the true mean has a maximum error of $d$. 

According to these observations, in order to obtain \eqref{eq:theocons}, it is necessary to examine the statistical concentration properties of the risk and of the empirical training mean. This task is complicated by the fact that both quantities depend on random functions $\bm{f}_k^N$. For this reason, we must resort to {\em uniform} (w.r.t. the class of functions) laws of large numbers. These types of concentration results are based notably on McDiarmid's inequality~\cite{mcdiarmid1998concentration}. \qed

In what concerns the expression in \eqref{eq:theocons}, the term $\rho_N^{(k)}$ contains the complexity of the chosen classifier structure, which depends itself on the training set size $N_k$. We will see in the next section that it evolves as $\mathcal{O}(1/\sqrt{N_k})$ for feedforward neural networks. Therefore as $N_k$ grows, we can neglect $\rho_N^{(k)}$. Now, since by assumption $d\in (0, -\log(e^{R(f^o)}-1))$, both terms $d$ and $\Delta-R(f^o)$ are strictly positive. This implies that both exponential terms in \eqref{eq:theocons} vanish, which in turn implies that the probability of consistent learning for the proposed strategy approaches $1$, as the training sets grow.

\subsection{Neural Network Complexity}
{In this section, we complement the result from Theorem~\ref{the:consist} by showing that the term $\rho_N^{(k)}$ in \eqref{eq:rho}, which depends on the Rademacher complexity of the classifier, vanishes with an increasing number of training samples in the case of feedforward neural networks (FNN).} Assume that one classifier has the structure of a FNN with $L$ layers (excluding the input layer) and activation function $\sigma$. We drop index $k$ as we are referring to a single FNN. Each layer $\ell$ consists of $n_\ell$ nodes, equivalently the size of layer $\ell$ is given by $n_\ell$. 

At each node $m=1,2,\dots,n_\ell$ of layers $\ell=2,3,\dots,L$, the following function $g_m^{(\ell)}$ is implemented:
\begin{equation}
g_m^{(\ell)}(h)=\sum_{j=1}^{n_{\ell-1}}w^{(\ell)}_{mj}\sigma\left(g^{(\ell-1)}_j(h)\right) -\theta^{(\ell)}_m.\label{eq:ffnn1}\vspace{-1pt}
\end{equation}
The parameters $w^{(\ell)}_{mj}$ correspond to the elements of the weight matrix $W_\ell$ of dimension $n_{\ell}\times n_{\ell-1}$. The offset parameters $\theta^{(\ell)}_m$ are the elements of a vector $\theta^{(\ell)}$ of dimension $n_{\ell}$. For the first layer, the function implemented at node $m$ is of the form:
\begin{equation}
g_m^{(1)}(h)=\sum_{j=1}^{n_0}w^{(1)}_{mj}h_j-\theta^{(1)}_m,\label{eq:ffnn2}\vspace{-1pt}
\end{equation}
where the input vector $h$ has dimension $n_0$. 

{For a FNN whose purpose is to solve a binary classification problem, we denote the output at layer $L$ by $z\in\mathbb{R}^2$, where  $z_m=g_m^{(L)}(h)$ for $m=1,2$. The final output is given by applying the softmax function to $z$. In this case the logit function is given by:
\begin{equation}
f^{\sf NN}(h)=\log\frac{ \widehat{p}(+1|h)}{\widehat{p}(-1|h)}=z_1-z_2\label{eq:ffnn3}
\end{equation}
where we say that $f^{\sf NN}$ belongs to a class of functions $\mathcal{F}^{\sf NN}$, which is parameterized by matrices $W_\ell$ and bias vectors $\theta^{(\ell)}$, for $\ell=1,2,\dots,L$, according to \eqref{eq:ffnn1}, \eqref{eq:ffnn2} and $\eqref{eq:ffnn3}$.

We are interested in finding an expression for the Rademacher complexity of class $\mathcal{F}^{\sf NN}$ described above. An upper bound for this complexity can be found in Lemma~\ref{lem:radnn} inspired by results from \cite{bartlett2002rademacher} (proof is omitted due to space limitations).}
\begin{lemma}[\textbf{Rademacher Complexity of FNNs}] \label{lem:radnn}Consider an $L$-layered feedforward neural network, satisfying $\|w^{(\ell)}_m\|_1\leq b$, $|\theta^{(\ell)}(m)|\leq a$, for every node $m=1,2,\dots, n_{\ell}$ and every layer $\ell=1,2,\dots,L$. Assume that the input vector $h\in\mathbb{R}^{n_0}$ satisfies $\|h\|_\infty\leq c$\footnote{$\|x\|_\infty$ denotes the $\ell_\infty$-norm defined as $\|x\|_\infty\triangleq \max_i |x_i|$.}, that the activation function $\sigma(x)$ is Lipschitz with constant $L_\sigma$ and that $\sigma(0)=0$. Then the Rademacher average for the set of vectors $\mathcal{F}^{\sf NN}(h_1^N)$ is bounded by:
	\begin{align}
	& \mathcal{R}(\mathcal{F}^{\sf NN}(h_1^N))\hspace{-2pt}\leq\hspace{-2pt} \frac{2}{\sqrt{N}}\hspace{-2pt}\left[(2bL_{\sigma})^{L-1}bc\sqrt{2\log(2n_0)}\hspace{-1pt}+\hspace{-2pt}\sum_{\ell=0}^{L-1}(2bL_{\sigma})^\ell a\right]\hspace{-2pt}.
	\end{align}\QED
\end{lemma}
\vspace{-20pt}
\section{Simulation Results}
\vspace{-5pt}
To illustrate the proposed strategy, we consider a network of $10$ agents, whose topology can be seen in Fig.~\ref{fig:net}. The combination matrix is generated using an averaging rule~\cite{sayed2014adaptation}, and we ensure that at least one agent possesses a self-loop. 

We consider the MNIST dataset \cite{lecun2010mnist}, using digits $0$ and $1$ for a binary classification task. Feature vectors are the $784$ pixels of each image of the handwritten digits. Each agent disposes of $98$ training samples for each class of digits. With this dataset, each agent trains its own classifier, which has the structure of a feedforward neural network with one hidden layer with $64$ nodes, and activation function $\text{arctan}(\cdot)$. To illustrate the robustness of the network of classifiers, we purposely tamper with the dataset for Agent $1$, highlighted in the {leftmost panel of }Fig.~\ref{fig:net}. To obtain a poor training performance, we provide agent $1$ with only digits $1$ during training and randomly assigned labels.

The training phase is run using mini-batch iterates of $10$ samples, over $15$ epochs. The empirical risk evolution at the individual agents as training progresses is shown in Fig.~\ref{fig:net} {(middle panel)}, where we can see how the training performance of agent $1$ is much worse than the performance of other agents. The average empirical risk, which is given by $\sum_{k=1}^K\pi_k\widetilde{\bm{R}}^N(\widetilde{\bm{f}}^N)$, is hardly affected by the deviating behavior of agent $1$.

In the prediction phase, all agents are receiving streaming observations, i.e., images of digits. Agents are observing digits $0$ until the time instant $500$, from which they start observing digits $1$. In the {rightmost panel of} Fig.~\ref{fig:net}, we see the classification variable $\bm{\lambda}_{1,i}$ of agent 1 over time, showing that, although agent $1$ has a poorly trained model, it is able to learn consistently the true state. 

%

\bibliographystyle{IEEEbib}
\balance
\bibliography{ref}

\begin{thebibliography}{10}

\bibitem{jadbabaie2012non}
A.~Jadbabaie, P.~Molavi, A.~Sandroni, and A.~Tahbaz-Salehi,
\newblock ``Non-{B}ayesian social learning,''
\newblock {\em Games and Economic Behavior}, vol. 76, no. 1, pp. 210--225,
  2012.

\bibitem{krishnamurthy2013social}
V.~Krishnamurthy and H.~V. Poor,
\newblock ``Social learning and {B}ayesian games in multiagent signal
  processing: How do local and global decision makers interact?,''
\newblock {\em IEEE Signal Processing Magazine}, vol. 30, no. 3, pp. 43--57,
  2013.

\bibitem{nedic2017fast}
A.~Nedi{\'c}, A.~Olshevsky, and C.~A. Uribe,
\newblock ``Fast convergence rates for distributed non-{B}ayesian learning,''
\newblock {\em IEEE Transactions on Automatic Control}, vol. 62, no. 11, pp.
  5538--5553, 2017.

\bibitem{salami2017social}
H.~Salami, B.~Ying, and A.~H. Sayed,
\newblock ``Social learning over weakly connected graphs,''
\newblock {\em IEEE Transactions on Signal and Information Processing over
  Networks}, vol. 3, no. 2, pp. 222--238, 2017.

\bibitem{lalitha2018social}
A.~Lalitha, T.~Javidi, and A.~D. Sarwate,
\newblock ``Social learning and distributed hypothesis testing,''
\newblock {\em IEEE Transactions on Information Theory}, vol. 64, no. 9, pp.
  6161--6179, 2018.

\bibitem{matta2020interplay}
V.~Matta, V.~Bordignon, A.~Santos, and A.~H. Sayed,
\newblock ``Interplay between topology and social learning over weak graphs,''
\newblock {\em IEEE Open Journal of Signal Processing}, vol. 1, pp. 99--119,
  2020.

\bibitem{bordignon2020adaptive}
V.~Bordignon, V.~Matta, and A.~H. Sayed,
\newblock ``Adaptive social learning,''
\newblock {\em submitted for publication}, available at arXiv:2004.02494
  [cs.MA], 2020.

\bibitem{zhao2012learning}
X.~Zhao and A.~H. Sayed,
\newblock ``Learning over social networks via diffusion adaptation,''
\newblock in {\em Proc. Asilomar Conference on Signals, Systems and Computers
  (ASILOMAR)}, 2012, pp. 709--713.

\bibitem{hare2020non}
J.~Z. Hare, C.~A. Uribe, L.~Kaplan, and A.~Jadbabaie,
\newblock ``Non-{B}ayesian social learning with uncertain models,''
\newblock {\em IEEE Transactions on Signal Processing}, vol. 68, pp.
  4178--4193, 2020.

\bibitem{kittler1998combining}
J.~Kittler, M.~Hatef, R.~P.~W. Duin, and J.~Matas,
\newblock ``On combining classifiers,''
\newblock {\em IEEE transactions on pattern analysis and machine intelligence},
  vol. 20, no. 3, pp. 226--239, 1998.

\bibitem{breiman1996bagging}
L.~Breiman,
\newblock ``Bagging predictors,''
\newblock {\em Machine learning}, vol. 24, no. 2, pp. 123--140, 1996.

\bibitem{freund1997decision}
Y.~Freund and R.~E. Schapire,
\newblock ``A decision-theoretic generalization of on-line learning and an
  application to boosting,''
\newblock {\em Journal of computer and system sciences}, vol. 55, no. 1, pp.
  119--139, 1997.

\bibitem{tresp2000bayesian}
V.~Tresp,
\newblock ``A {B}ayesian committee machine,''
\newblock {\em Neural computation}, vol. 12, no. 11, pp. 2719--2741, 2000.

\bibitem{nguyen2008local}
D.~Nguyen-Tuong, J.~Peters, and M.~Seeger,
\newblock ``Local {G}aussian process regression for real time online model
  learning and control,''
\newblock in {\em Proc. International Conference on Neural Information
  Processing Systems}, 2008, pp. 1193--1200.

\bibitem{lederer2020real}
A.~Lederer, A.~J.~O. Conejo, K.~Maier, W.~Xiao, and S.~Hirche,
\newblock ``Real-time regression with dividing local {G}aussian processes,''
\newblock available at arXiv:2006.09446 [cs.LG], 2020.

\bibitem{matta2018estimation}
V.~Matta and A.~H. Sayed,
\newblock {\em Estimation and detection over adaptive networks}, pp. 69--106,
\newblock Elsevier, 2018.

\bibitem{bordignon2020adaptation}
V.~Bordignon, V.~Matta, and A.~H. Sayed,
\newblock ``Adaptation in online social learning,''
\newblock in {\em Proc. European Signal Processing Conference (EUSIPCO)}, 2020,
  pp. 2170--2174.

\bibitem{mohri2018foundations}
M.~Mohri, A.~Rostamizadeh, and A.~Talwalkar,
\newblock {\em Foundations of Machine Learning},
\newblock MIT press, 2018.

\bibitem{kay1993fundamentals}
Steven~M Kay,
\newblock {\em Fundamentals of Statistical Signal Processing: Detection
  Theory},
\newblock Prentice Hall PTR, 1993.

\bibitem{poor2013introduction}
H~Vincent Poor,
\newblock {\em An Introduction to Signal Detection and Estimation},
\newblock Springer Science \& Business Media, 2013.

\bibitem{sayed2014adaptation}
A.~H. Sayed,
\newblock ``Adaptation, learning, and optimization over networks,''
\newblock {\em Foundations and Trends in Machine Learning}, vol. 7, no.
  ARTICLE, pp. 311--801, 2014.

\bibitem{koltchinskii2001rademacher}
V.~Koltchinskii,
\newblock ``Rademacher penalties and structural risk minimization,''
\newblock {\em IEEE Transactions on Information Theory}, vol. 47, no. 5, pp.
  1902--1914, 2001.

\bibitem{boucheron2005theory}
S.~Boucheron, O.~Bousquet, and G.~Lugosi,
\newblock ``Theory of classification: A survey of some recent advances,''
\newblock {\em ESAIM: Probability and Statistics}, vol. 9, pp. 323--375, 2005.

\bibitem{mcdiarmid1998concentration}
C.~McDiarmid,
\newblock ``Concentration,''
\newblock in {\em Probabilistic methods for algorithmic discrete mathematics},
  pp. 195--248. Springer, 1998.

\bibitem{bartlett2002rademacher}
P.~L. Bartlett and S.~Mendelson,
\newblock ``Rademacher and gaussian complexities: Risk bounds and structural
  results,''
\newblock {\em Journal of Machine Learning Research}, vol. 3, no. Nov, pp.
  463--482, 2002.

\bibitem{lecun2010mnist}
Y.~LeCun, C.~Cortes, and C.~J. Burges,
\newblock ``Mnist handwritten digit database,'' 2010,
\newblock ATT Labs [Online]. Available: \url{http://yann.lecun.com/exdb/mnist}.

\end{thebibliography}

\end{document}